\documentclass[prb, twocolumn, superscriptaddress, longbibliography]{revtex4-2}

\usepackage{
amsmath, 
amssymb, 
bbm, 
bm, 
braket, 
color, 
graphicx, 
amsfonts, 
dsfont, 
xfrac
}

\usepackage[colorlinks, citecolor=blue, urlcolor=blue]{hyperref}

\usepackage[normalem]{ulem}

\newcommand{\pd}{{\phantom{\dag}}}

\begin{document}

\title{Real-space topological localizer index to fully characterize the dislocation skin effect}

\author{Nisarg Chadha}
\affiliation{Undergraduate Programme, Indian Institute of Science, Bangalore 560012, India}

\author{Ali G. Moghaddam}
\affiliation{Computational Physics Laboratory, Physics Unit, Faculty of Engineering and Natural Sciences, Tampere University, FI-33014 Tampere, Finland}
\affiliation{Department of Physics, Institute for Advanced Studies in Basic Sciences (IASBS), Zanjan 45137-66731, Iran}
\author{Jeroen van den Brink}
\affiliation{Leibniz Institute for Solid State and Materials Research, IFW Dresden, Helmholtzstraße 20, 01069 Dresden, Germany}
\affiliation{Institute for Theoretical Physics,
Technische Universit\"{a}t Dresden, 01069 Dresden, Germany}
\affiliation{W\"{u}rzburg-Dresden Cluster of Excellence ct.qmat}

\author{Cosma Fulga}
\affiliation{Leibniz Institute for Solid State and Materials Research, IFW Dresden, Helmholtzstraße 20, 01069 Dresden, Germany}
\affiliation{W\"{u}rzburg-Dresden Cluster of Excellence ct.qmat}

\date{\today}

\begin{abstract}
The dislocation skin effect exhibits the capacity of topological defects to trap an extensive number of modes in two-dimensional non-Hermitian systems.
Similar to the corresponding skin effects caused by system boundaries, this phenomenon also originates from nontrivial topology. However, finding the relationship between the dislocation skin effect and nonzero topological invariants, especially in disordered systems, can be obscure and challenging. Here, we introduce a real-space topological invariant based on the {\it spectral localizer} to characterize the skin effect on two-dimensional lattices. We demonstrate that this invariant consistently predicts the occurrence and location of both boundary and dislocation skin effects, offering a unified approach applicable to both ordered and disordered systems. Our work demonstrates a general approach that can be utilized to diagnose the topological nature of various types of skin effects, particularly in the absence of translational symmetry when momentum-space descriptions are inapplicable.
\end{abstract}

\maketitle

\section{Introduction}
The global conservation of energy ensures that the dynamics of the system together with its environment is Hermitian. 
However, in some cases it is more convenient to treat the system separately, while introducing the external coupling effectively as a non-Hermitian interaction in the system~\cite{ashida2020non}. 
Non-Hermitian descriptions are thus commonly used to study optical systems with gain and loss~\cite{wang2021topological}, electronic circuits with external contacts~\cite{datta2005quantum}, atomic systems coupled to probes~\cite{PhysRevLett.124.250402}, or acoustic systems~\cite{zhang2021acoustic}. 
In these cases, energies can be complex, and eigenstates are not guaranteed to form an orthonormal basis~\cite{brody2013biorthogonal}, leading to phenomena that have no counterpart in Hermitian systems.

One such phenomenon is the non-Hermitian skin effect (NHSE)~\cite{PhysRevLett.121.086803, PhysRevLett.124.086801, RevModPhys.93.015005, zhang2022review, Banerjee_2023, lin2023topological}, which denotes the localisation of an extensive number of eigenstates at the boundary of the system~\cite{PhysRevB.97.121401}. 
The NHSE is a consequence of nontrivial bulk topology: 
With periodic boundary conditions, the nonzero winding number of the bulk spectrum around a point in the complex plane marks the presence of a nontrivial point gap within which boundary states accumulate~\cite{PhysRevLett.124.056802, PhysRevLett.124.086801}.

Recently it has been pointed out that the NHSE is not necessarily a boundary property, but that it may also occur at topological defects such as dislocations \cite{PhysRevB.104.L161106, PhysRevB.104.L241402, PhysRevB.106.L041302} and disclinations \cite{PhysRevLett.127.066401}.
In this regard, the topology of non-Hermitian systems parallels that of Hermitian ones, allowing the application of conventional bulk-defect correspondence \cite{PhysRevB.82.115120, RevModPhys.88.035005, teo2017topological} to determine the combinations of system symmetries and defect types that lead to topologically protected gapless modes. In practice, however, there are several factors that complicate the task of computing the topological invariants responsible for the defect NHSE.
For example, previous works have shown examples of systems where dislocations host a NHSE for which the conventional bulk-defect correspondence does not apply~\cite{PhysRevB.104.L161106, PhysRevB.104.L241402}.
Moreover, the topological invariants are usually computed in an effective Brillouin zone composed of the original momentum space and supplemented by additional degrees of freedom which parameterize the surface surrounding the defect~\cite{PhysRevB.82.115120, teo2017topological}.
It is not a priori clear how this can be done when momentum is not a good quantum number, as is the case in disordered, fractal, quasicrystalline, or even amorphous models~\cite{manna2023inner}.

In this work, we examine the defect-induced NHSE from a different perspective.
We turn to a real-space topological invariant called the \emph{localizer index}.
The latter is one of a family of versatile topological invariants that were initially introduced to study Hermitian topological insulators~\cite{loring2015k, PhysRevLett.116.257002, grossmann2016index, loring2017finite, 10.1063/1.5094300, 10.1063/1.5083051, loring2020spectral, doll2020approximate, DOLL2021108038}, but have since been extended to study a variety of phases.
These include metals and semimetals~\cite{schulz2022invariants, PhysRevB.106.064109,cheng2023revealing,franca2023obstructions,franca2023topological}, higher-order topological phases~\cite{PhysRevB.106.064109}, and more recently the 1D NHSE~\cite{ochkan2023observation}, Floquet phases ~\cite{PhysRevB.108.035107}, as well as line-gapped non-Hermitian phases~\cite{cerjan2023spectral}.
We show that one particular localizer index, originally meant to characterize one-dimensional (1D) Hermitian systems~\cite{loring2015k}, can be adapted to study the topological properties of the NHSE in two-dimensional (2D), point-gapped non-Hermitian systems.
One of its advantages is that, given a concrete system, it allows for the direct detection of the topology associated to both boundaries as well as dislocations.
This approach sidesteps the need for constructing an effective Brillouin zone, and is thus ideally suited for the study of disordered systems.

We begin in Sec.~\ref{sec:model} by introducing a simple model of a topologically-nontrivial 2D non-Hermitian system, constructed as a stack of parallel Hatano-Nelson chains~\cite{PhysRevB.104.L161106,PhysRevB.104.L241402}.
We describe the process of introducing different types of dislocations into the system and show that they host a NHSE, similar to the boundaries of the model.
Sec.~\ref{sec:topo_defect} is devoted to understanding the observed NHSEs from a topological point of view, based on a mapping that relates the topology of Hermitian and non-Hermitian systems.
After a brief review of previous approaches, we introduce the spectral localizer, describe its application to 1D Hermitian models, and expand its usage to 2D non-Hermitian models.
We highlight the advantages of a direct, real-space formulation of the topological invariant in Sec.~\ref{sec:disorder}, by showing that it correctly predicts the robustness of the NHSE against onsite potential disorder.
Finally, we conclude in Sec.~\ref{sec:summary}, suggesting that a variety of different types of skin effects, in systems of varying dimensionality and symmetry class, may be amenable to a localizer-based topological description.

\section{\label{sec:model}Model}

The Hatano-Nelson (HN) model is one of the simplest systems exhibiting the NHSE~\cite{PhysRevLett.77.570, PhysRevB.56.8651}.
It consists of a 1D chain with one orbital per unit cell and nearest neighbour non-reciprocal hoppings given by $t_x(1\pm\gamma)$.
Under open boundaries, the Hatano-Nelson chain shows the NHSE with an exponential accumulation of all eigenstates towards the boundary.
The direction of accumulation is given by the largest hopping.

Following Ref.~\cite{PhysRevB.104.L241402} we create a periodic 2D system by stacking HN chains with an inter-chain coupling strength $t_y$.
This gives the weak Hatano-Nelson model, with the Hamiltonian 
\begin{equation}\label{eq:weakHN}
 H(\mathbf{k})=2t_x\cos{k_x}+2t_y\cos{k_y}-2i\gamma t_x\sin{k_x},
\end{equation}
where $\mathbf{k}=(k_x, k_y)$ the momentum vector.
In the following we will set $t_x=1$ as the energy scale of the problem, expressing all other energy scales relative to it.
All numerical results are obtained using the Kwant library \cite{Groth2014} and our own code is included in the Supplemental Material.

\begin{figure}[tb]
\includegraphics[width=0.99\columnwidth]{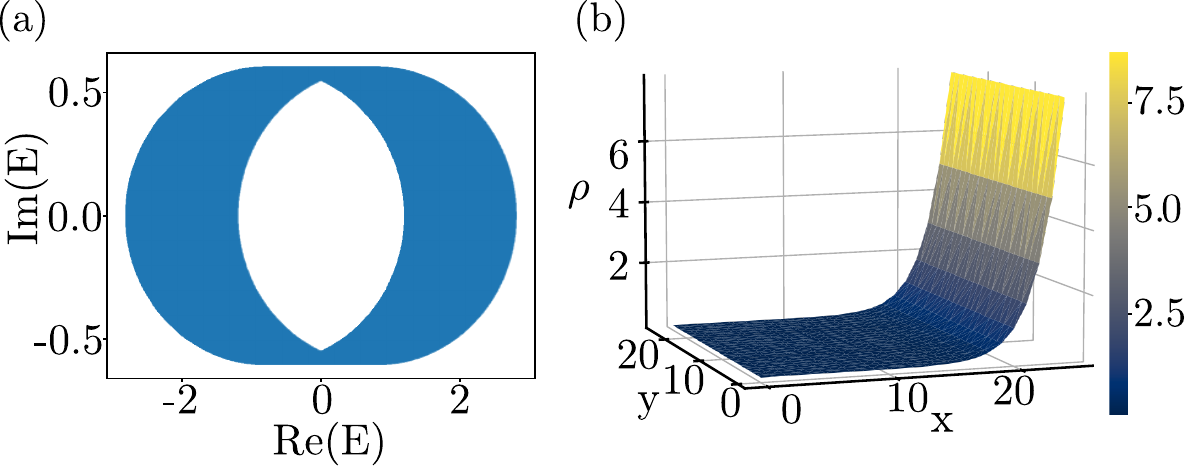}
\caption{
Panel (a) shows the spectrum of the momentum-space Hamiltonian Eq.~\eqref{eq:weakHN} in the complex energy plane.
The spectrum consists of a set of ellipses displaced relative to each other along the real axis, and shows a point gap around the origin, $E=0$.
Panel (b) shows the SPD of Eq.~\eqref{eq:SPD} for a finite-sized system consisting of 25$\times$25 sites with OBC.
For ease of visualization, $\rho$ is also shown as a varying color scale.
For both panels, we use $t_y=0.4$ and $\gamma=0.3$ in units of $t_x$.
}
\label{fig:Stack}
\end{figure}

For $|t_y|<1$, the complex spectrum has a point-gap [shown in Fig.~\ref{fig:Stack}(a)]. It can be considered as a collection of periodic HN chains with a momentum-dependent chemical potential $2t_y\cos{k_y}$ \cite{PhysRevB.104.L241402}, such that the spectrum consists of a set of ellipses displaced relative to each other along the real energy axis.
For a finite-sized system with open boundary conditions (OBC), the non-reciprocity along the $x$-direction leads to the formation of a NHSE.
To describe the latter, we turn to the real-space probability density summed over all states,
\begin{equation}\label{eq:SPD}
\rho(\mathbf{r})=\sum_{n}|\braket{\mathbf{r}|\psi_n}|^2,
\end{equation}
where $\mathbf{r}$ is the position of a lattice site, $\ket{\mathbf{r}}$ is the position ket, $\ket{\psi_n}$ is the $n^{\rm th}$ right eigenstate of the Hamiltonian, and the sum runs over all states.
The summed probability density (SPD) $\rho$, plotted in Fig.~\ref{fig:Stack}(b), shows an exponential accumulation towards the boundary.
In effect, each open HN chain in the stack produces its own non-Hermitian skin effect, with the same, $y$-independent localization length.

\begin{figure}[tb]
\includegraphics[width=0.99\columnwidth]{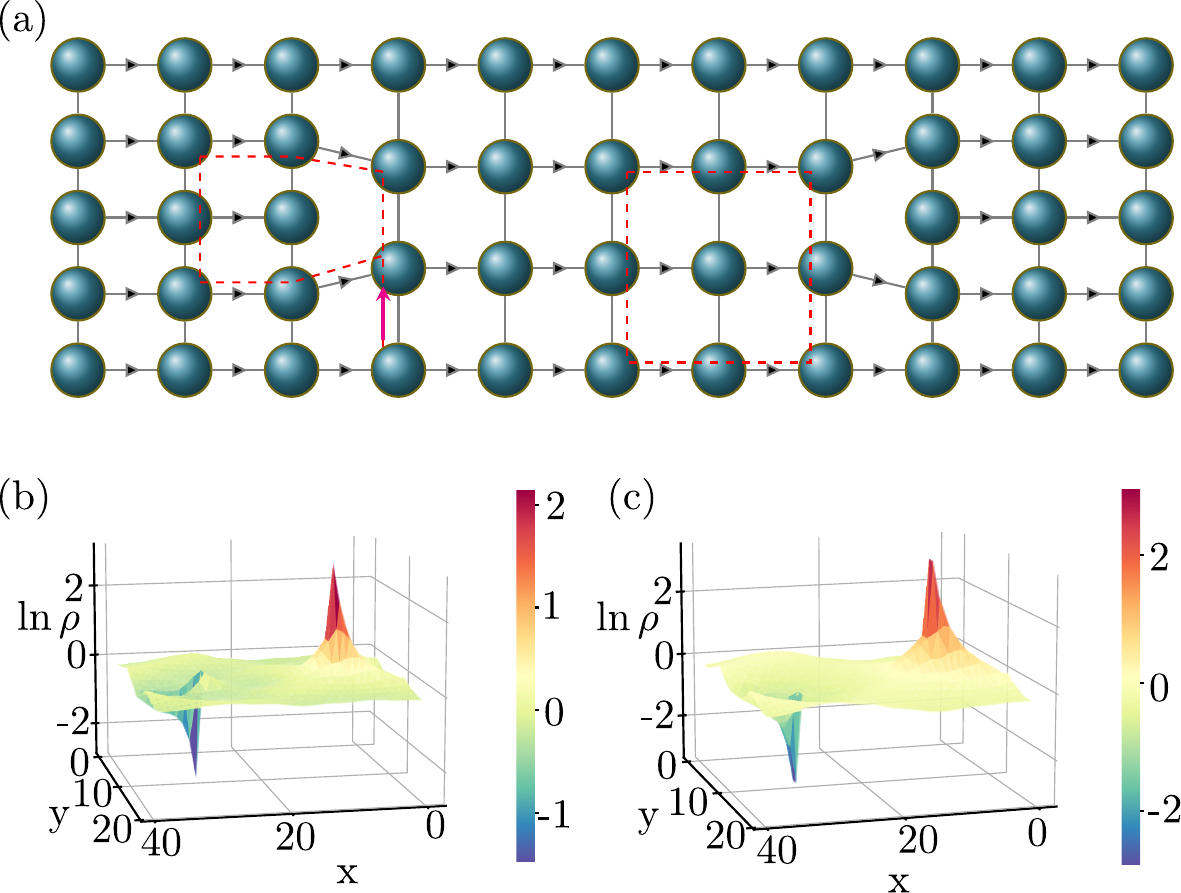}
\caption{Panel (a) shows a sketch of the the weak Hatano-Nelson model in the presence of two dislocations. The dashed lines show two closed contours, one of which encircles the dislocation and one which does not.
The path that encircles the defect ends up with a net displacement equal to the Burgers vector, here $\mathbf{B}=(B_x, B_y) = (0, 1)$, shown by the thick arrow. Panels (b, c) show the skin and anti-skin effect at the two dislocations. 
There is an accumulation or a depletion of $\rho$ compared to its bulk value depending on whether $B_y$ is positive or negative.
In each case, the system size is $40\times20$ sites, the distance between the two dislocations is 20 sites, $t_y=0.4$, and $\gamma=0.4$.
Panel (b) shows the case of unit Burgers vectors, $\mathbf{B}=(0, \pm 1)$, whereas the dislocations in panel (c) have $\mathbf{B}=(0, \pm 2)$.
}
\label{fig:dislocation}
\end{figure}

We introduce dislocations in the lattice by removing one or more rows of sites at fixed $y$-coordinates and gluing the two resulting edges together using the same $t_y$ hopping as in the rest of the bulk.
An example of a system formed in this way is shown in Fig.~\ref{fig:dislocation}(a), and contains two dislocations.
Each is characterized by a Burgers vector, the additional translation required to form a closed loop around the dislocation core, compared to a loop that does not encircle the defect~\cite{chaikin1995principles, teo2017topological}.
In units of the lattice constant, the left-most dislocation has a Burgers vector $\mathbf{B}=(B_x, B_y)=(0, 1)$, whereas $\mathbf{B}=(0, -1)$ for the right-most dislocation.

We use periodic boundary conditions (PBC) in order to suppress the boundary NHSE, and examine the effect of the dislocations on the SPD in Fig.~\ref{fig:dislocation}(b).
As a result, we reproduce the findings of Ref.~\cite{PhysRevB.104.L241402}:
Depending on the sign of $B_y$, there is either an accumulation or a depletion of the density relative to that far from the dislocations.
These phenomena have been termed the skin and anti-skin effect.

Going beyond the previous results of Ref.~\cite{PhysRevB.104.L241402}, we also turn to a system with double the Burgers vectors. This is achieved by removing two rows of sites in the cut and glue procedure, yielding $B_y=\pm 2$.
The SPD, shown in Fig.~\ref{fig:dislocation}(c), shows the peak and dip at the point defects in the same manner as for $B_y=\pm1$, but with a larger amplitude and width.
Thus, the NHSE is still present regardless of the parity of $B_y$.
Finally, we note that with OBC the boundary NHSE hinders the visibility of the dislocation NHSE.
We find that the peak and dip appearing in Fig.~\ref{fig:dislocation}(b, c) are no longer visible in this case, except in the regime of weakly-coupled chains, $|t_y|\ll 1$.

\section{\label{sec:topo_defect} Topology of the dislocation skin effect}

The topological properties of non-Hermitian systems with a point gap can be studied by means of a \emph{Hamiltonian-doubling procedure}, which maps them to Hermitian systems with the same topological classifications~\cite{PhysRevLett.124.086801}. 
Specifically, for a non-Hermitian Hamiltonian $H$ we construct a Hermitian
\begin{equation}
\label{eq:chiral_doubled}
\widetilde{H}=
\begin{pmatrix}
0&H\\
H^{\dagger}&0
\end{pmatrix}.
\end{equation}
The latter obeys chiral symmetry, $\Gamma \widetilde{H} = - \widetilde{H} \Gamma$, with $\Gamma={\rm diag}(\mathbb{I}, -\mathbb{I})$, where $\mathbb{I}$ is an identity matrix of the same size as $H$.

The NHSE present in the non-Hermitian $H$ maps to the topologically-protected zero energy modes of the Hermitian $\widetilde{H}$, and the two systems have equal-valued topological invariants~\cite{PhysRevLett.124.086801}.
For the Hatano-Nelson Hamiltonian $H$, the doubled Hamiltonian $\widetilde{H}$ is an SSH chain~\cite{PhysRevLett.42.1698,asboth2016short} whose 1D winding number is the same as the point-gap winding number of $H$.
Therefore, the topological invariant describing the SPD accumulation in the Hatano-Nelson model is the same as the bulk winding number for the doubled system.

\begin{figure}[tb]
\includegraphics[width=0.99\columnwidth]{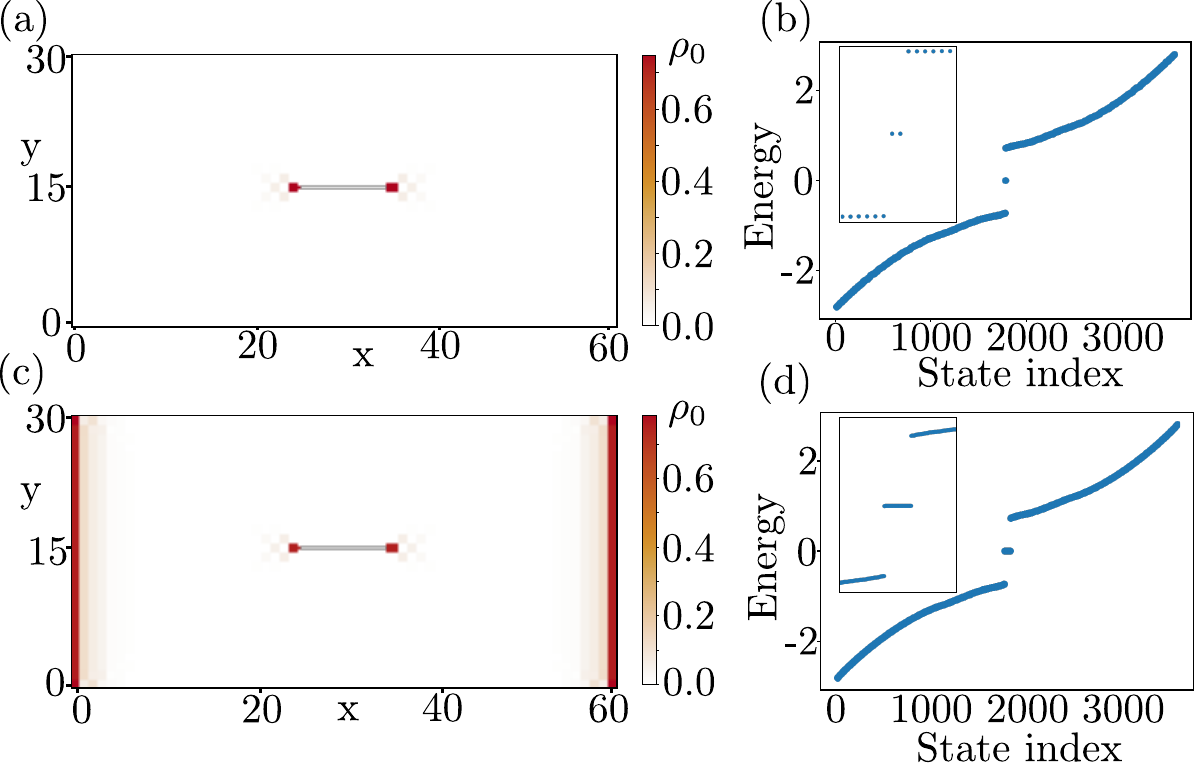}
\caption{
The local density of zero modes, $\rho_0(\mathbf{r})$, for the doubled Hamiltonian Eq.~\eqref{eq:chiral_doubled} is shown in panel (a) for PBC and in panel (c) for OBC. $\rho_0(\mathbf{r})$ is the summation of the eigenvector probability carried out over only the zero energy modes, defined using a tolerance of $10^{-4}$ in units of $t_x$.
Correspondingly, the spectra of the two systems are shown in panels (b) and (d).
In all plots, we use $t_y=\gamma=0.4$, a system size of $60\times30$ sites, and introduce dislocations with $B_y=\pm 1$ that are 10 sites apart.
In the case of PBC, there are only two zero-energy modes, each one localized at a dislocation core.
With OBC, these defect modes coexist with boundary states formed by the topological end-modes of each SSH chain in the stack.
}
\label{fig:obcpbc}
\end{figure}

For our case, the doubling procedure Eq.~\eqref{eq:chiral_doubled} maps the stack of Hatano-Nelson chains into a stack of SSH chains.
At each dislocation, we observe one zero-energy state in the case $|B_y|=1$, whereas two states are present at each defect when $|B_y|=2$.
We note that when the doubled system has OBC, the zero modes at the bulk defects coexist with gapless states at the boundaries of the system, as shown in Fig.~\ref{fig:obcpbc}.
Thus, we expect that the topological invariants characterizing the dislocations should be identifiable under OBC, even though the boundary NHSE, when present, obscures the presence of a defect NHSE.

\subsection{Previous approaches to characterize the dislocation skin effect}

According to the conventional bulk-defect correspondence established for Hermitian systems \cite{PhysRevB.82.115120}, the topological invariant characterizing a dislocation in 2D is computed from a three-dimensional (3D) effective Hamiltonian that surrounds the defect, $\widetilde{H}(k_x, k_y, s)$, where $k_{x,y}$ are the two original bulk momenta, and $s\in[0,1]$ is a periodic variable describing a circle around the defect.
For a non-Hermitian $H$ that does not have any additional symmetries (class A in the Altland-Zirnbauer classification~\cite{PhysRevB.55.1142, PhysRevX.9.041015}), $\widetilde{H}$ belongs to class AIII, and the dislocation invariant is expected to take the form of a 3D winding number:
\begin{equation}
    \label{eq:w3}
    W_3=\int_{{\rm BZ}\times \mathcal{S}} \frac{d^2\mathbf{k}ds}{12\pi}\epsilon_{\mu\nu\rho} {\rm Tr}[(q^{-1}\partial_{\mu}q)(q^{-1}\partial_{\nu}q)(q^{-1}\partial_{\rho}q)],
\end{equation}
where $\epsilon_{\mu\nu\rho}$ is the anti-symmetric Levi-Civita tensor, and $q(\mathbf{k},s)$ is the off-diagonal block of the Hermitian Hamiltonian $\widetilde{H}(k_x, k_y, s)$ in a basis where the chiral symmetry operator is of the form ${\rm diag}(\mathbb{I}, -\mathbb{I})$.
Using the doubling construction Eq.~\eqref{eq:chiral_doubled} means that for the stack of Hatano-Nelson chains $q=H$.
As pointed out in Ref.~\cite{PhysRevB.104.L241402}, however, since the model has a single band, $q(\mathbf{k},s)$ is a scalar, the Levi-Civita summation gives a vanishing contribution to the invariant.
Therefore, $W_3$ fails to capture the topology of the one-band model, regardless of which type of dislocation is considered.

As an alternative to the 3D winding number, Ref.~\cite{PhysRevB.104.L161106} proposed an invariant given by the 1D winding number of the bulk Hamiltonian along specific lines of the Brillouin zone for which ${\bf B}\cdot {\bf k} = \pi\, {\rm mod}\, 2\pi$. 
Thus, for $B_y=\pm 1$, the index is the $k_x$ winding number at $k_y=\pi$, while for $B_y=\pm 2$ it is the sum of $k_x$ winding numbers at $k_y=\pm\pi/2$. 
Ref.~\cite{PhysRevB.104.L241402}, on the other hand, proposed an invariant of the form $\vartheta = \nu_x B_y - \nu_y B_x$, where $\nu_x$ and $\nu_y$ are weak topological invariants that predict the appearance of the boundary NHSE.
These indices are defined as the 1D winding numbers of the bulk spectrum along a particular momentum direction, averaged over the perpendicular momentum direction.
Thus,
\begin{eqnarray}
\nu_j = \int \frac{d^2{\bf k}}{i(2\pi)^2}\: {H}({\bf k})^{-1} \,\partial^{\phantom\dag}_{k_j}{H}({\bf k})^{\phantom\dag} 
\label{eq:average-windings}, 
\end{eqnarray}
with $j=x,y$.
Here $\vartheta$ can take arbitrary integer values, consistent with the observation of a NHSE both for $B_y=\pm 1$ and $\pm 2$.
It can be derived starting from a Chern-Simons invariant defined in the effective Brillouin zone $(k_x, k_y, s)$ \cite{PhysRevB.104.L241402}, but the latter only captures the parity of $\vartheta$ and does not yield the expected $\mathbb{Z}$ classification.
And the fact that both Refs.~\cite{PhysRevB.104.L161106, PhysRevB.104.L241402} define invariants in momentum space hinders their use in disordered systems.

\subsection{Localizer index}

We turn to a real space description of the dislocation NHSE.
To this end, we consider a Hermitian matrix called the \emph{spectral localizer} ~\cite{loring2015k, loring2019guide}, which is constructed from a real-space, 1D Hermitian Hamiltonian in class AIII. It takes the form
\begin{equation}
\label{eq:localizer}
L = (\widetilde{X}+i\widetilde{H})\Gamma,
\end{equation}
where $\widetilde{H}$ is the Hamiltonian matrix, $\Gamma$ is the chiral symmetry matrix, and $\widetilde{X}={\rm diag} ( x_1-x_0, x_2-x_0, x_3-x_0, \ldots)$ contains the positions of the lattice sites relative to a given origin, $x_0$.
From $L$, it is possible to define a $\mathbb{Z}$-valued topological invariant called localizer index~\cite{loring2015k}:
\begin{eqnarray}\label{eq:localizer_herm}
\nu_L = \frac{1}{2} {\rm Sig} \, L,
\end{eqnarray}
where $\rm{Sig}$ refers to the matrix signature --
the difference in the number of positive and negative eigenvalues. 

For an SSH chain with OBC, $\nu_L$ predicts the number of zero-energy modes at each end whenever the origin of space, $x_0$, is positioned deep in the bulk of the chain.
It gives a trivial answer when the origin is outside of the chain, e.g. when all lattice site positions $x_j>x_0$.
In effect, the localizer index is equal to the net number of zero-energy modes (counted with their chirality), at positions away from the origin $x_0$.

In our case, the Hamiltonian doubling procedure Eq.~\eqref{eq:chiral_doubled} yields an array of SSH chains oriented along the $x$-direction and stacked along $y$.
As shown in Fig.~\ref{fig:obcpbc}, with OBC $\widetilde{H}$ hosts zero modes both at its boundaries, in correspondence to the boundary NHSE, as well as at dislocations, in correspondence to the dislocation NSHE.
Thus, we expect that the localizer index will give a unified prediction for both boundary as well as defect states by considering $\nu_L$ as a function of $x_0$.

\begin{figure}[tb]
\includegraphics[width=0.99\columnwidth]{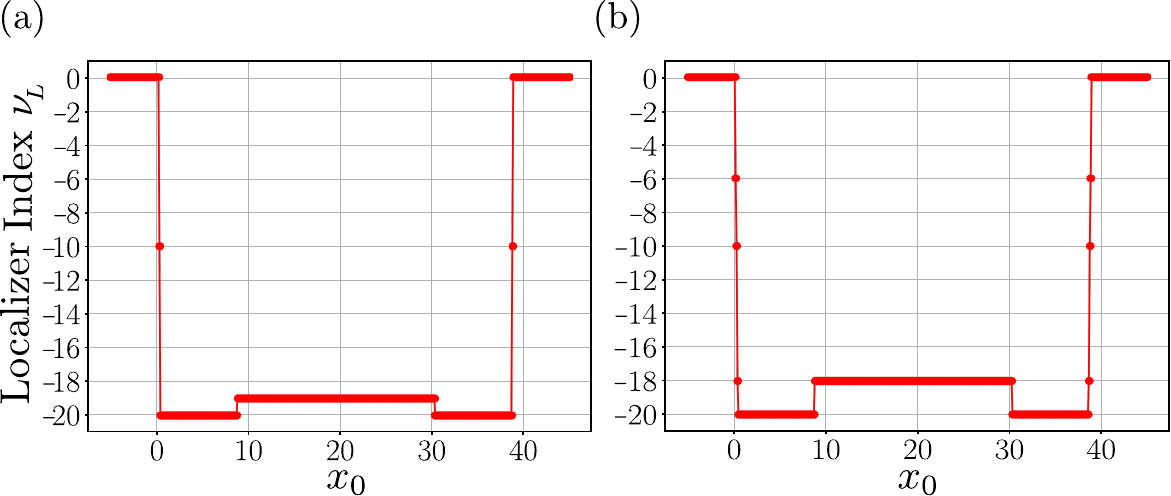}
\caption{
Panel (a) shows the topological index sweeping over $x_0$ for a $40\times 20$ lattice with $B_y=\pm1$, $\gamma=0.4, t_y=0.4$. The index shows a jump equal to 1 when the origin is between the point defects. Panel (b) shows the variation of $\nu_L$ for the same system with $B_y=\pm2$. $\nu_L$ jumps by 2 between the point-defects, showing the complete topological classification yielded by the localizer index.
}
\label{fig:B_2}
\end{figure}

We have computed the localizer index for the doubled Hamiltonian Eq.~\eqref{eq:chiral_doubled}, taking $\widetilde{X}$ to represent the lattice positions in the horizontal $x$ direction and independent of the sites' position in the vertical $y$ direction.
As shown in Fig.~\ref{fig:B_2}, $\nu_L$ correctly captures the number of zero modes as $x_0$ is varied across the system, both in the case $B_y=\pm 1$ as well as for $B_y=\pm 2$.
When $x_0$ is positioned outside the lattice, $\nu_L=0$ yields a trivial answer.
As $x_0$ enters the bulk of the system, $\nu_L$ changes by an amount equal to the number of boundary zero-modes, i.e. the number of SSH chains.
The same invariant, however, also correctly identifies the number of zero-modes bound to the dislocation, showing a jump by $\pm 1$ in Fig.~\ref{fig:B_2}(a) or by $\pm 2$ in Fig.~\ref{fig:B_2}(b) as $x_0$ is swept across the dislocation core.
Thus, the localizer index also allows to determine the position of the topologically protected states:
A difference of $\nu_L$ between two different values of $x_0$ is a topological invariant counting the number of protected zero-modes in a particular region of space.

Due to the mapping between the NHSE of the weak Hatano-Nelson model $H$ and the zero modes of the stack of SSH chain $\widetilde{H}$, the localizer index can thus predict the appearance and position of the NHSE both at boundaries as well as at dislocations.

We now go one step further and re-express the localizer index in such a way that it depends on the non-Hermitian Hamiltonian directly, thus avoiding the need for a doubling procedure.
As detailed in Appendix~\ref{app:Sylvester}, using the block LDU decomposition together with Sylvester's law of inertia we obtain
\begin{equation}
    \label{eq:nhlocalier}
    \nu_L=\frac{1}{2}{\rm Sig}(X)-\frac{1}{2}{\rm Sig} \left( X+HX^{-1}H^{\dagger} \right),
\end{equation}
where $X$ is the position operator corresponding to the non-Hermitian $H$.
In order to prevent $X$ from becoming singular, this formula requires that the origin of space $x_0$ be chosen such that it does not exactly coincide with one of the lattice site positions, which can be achieved for any finite discrete system.
Importantly, it is faster to evaluate numerically:
For a matrix of size $n\times n$, computing the signature has ${\cal O}(n^3)$ complexity, which means that Eq.~\eqref{eq:nhlocalier} is eight times faster than Eq.~\eqref{eq:localizer_herm}, while producing, as we have checked, an answer identical to that shown in Fig.~\ref{fig:NH_Index}.

We note that the localizer index' ability to describe the topological properties of both boundaries as well as dislocations in a unified manner goes beyond the conventional bulk-boundary and bulk-defect correspondence. 
According to the latter, different invariants are generally needed for the two, as mentioned at the beginning of this section.

\section{\label{sec:disorder}Robustness against disorder}

The NHSE occuring in the weak Hatano-Nelson model does not require any symmetry, such that it is expected to be robust against disorder.
We test this hypothesis by adding on-site potential disorder to the model, choosing for each site $j$ a random potential $\omega_j$, drawn independently from the uniform distribution $[-W/2,W/2]$.
$W$ therefore encodes the strength of disorder.

\begin{figure}[t]
\includegraphics[width=0.99\columnwidth]{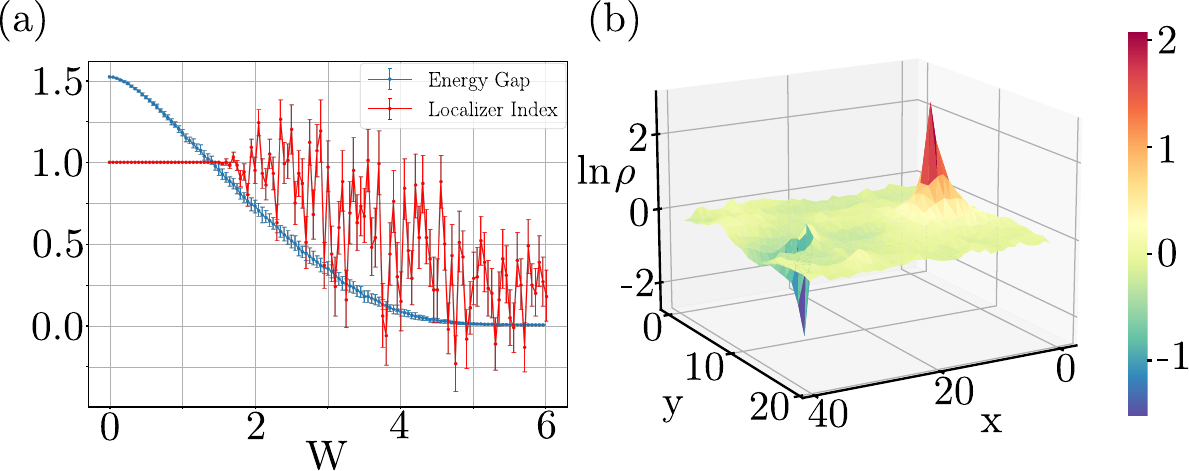}
\caption{
Panel (a): Average localizer index characterizing the dislocation NHSE (red), and average bulk gap of the doubled Hamiltonian (blue) as a function of disorder strength $W$.
The system consists of $40\times 20$ lattice sites and contains two dislocations with $B_y=\pm 1$ that are positioned 20 lattice sites apart. We use $t_y=\gamma=0.4$ and each point is obtained by averaging over 100 independent disorder realizations.
The localizer index characterizing the dislocation topology is obtained as a difference between the values of $\nu_L$ computed at $x_0=20.1$ and $x_0=7.1$.
Panel (b) shows the SPD for a single disorder configuration at disorder strength $W=1$. The skin and anti-skin peaks can be prominently seen.
}
\label{fig:disorder}
\end{figure}

Beyond testing for the robustness of the dislocation NHSE, adding disorder also allows us to check the validity of the localizer index in a regime where previous invariants do not apply, since momentum is not a good quantum number.
We show in Fig.~\ref{fig:disorder}(a) the average topological invariant describing the defect NHSE for a system containing dislocations with $B_y=\pm 1$. The index is computed as the difference of $\nu_L$ [Eq.~\eqref{eq:nhlocalier}] for two values of $x_0$ on either size of the left-most dislocation ($x_0=7.1$ and $x_0=20.1$).
We compare it with another indicator of NHSE robustness, the bulk gap of the doubled Hamiltonian.
We find that the system remains robust against disorder up to values of $W$ of the order of $t_x$, showing a well-quantized average invariant.
When disorder strength is increased further, the bulk gap decreases and the index loses its quantization.
In Fig.~\ref{fig:disorder}(b) we examine the SPD of a single disorder configuration at $W=1$.
While $\rho$ is clearly noisy in the bulk of the system compared to Fig.~\ref{fig:dislocation}(b), the peak and dip corresponding to the skin and anti-skin effect are clearly visible, consistent with the well-quantized localizer index.
Finally, we note that neither the distribution of gap sizes nor that of $\nu_L$ is gaussian, so that the error bars of Fig.~\ref{fig:disorder} cannot be determined simply from the variance. We detail their calculation in Appendix \ref{app:errorbar}.

\section{\label{sec:summary}Summary and Outlook}

In this work, we have revisited the skin effect occurring at dislocations in 2D non-Hermitian systems, using the simple toy model introduced in Refs.~\cite{PhysRevB.104.L161106, PhysRevB.104.L241402}.
We have found that a real-space topological invariant, called the localizer index, can fully capture the presence and the position of the NHSE, both at boundaries as well as at dislocations.
This is in contrast to previous approaches, which rely on different invariant formulas.

One of the main advantages of localizer invariants is their ability to probe a given system directly and without the need for momentum space.
Thus, we expect that this index may play an especially useful role in those systems where momentum space is inaccessible, such as in disordered or in amorphous systems.

Finally, we note that the index we have used is one of a large family of localizer invariants, which have been shown to apply to a variety of different Hermitian topological insulators, with different symmetries as well as in different dimensions.
Due to the mapping relating the topology of Hermitian and non-Hermitian Hamiltonians, we expect that similar types of localizer index can be useful to characterize other types of NHSE, such as the ones protected by time-reversal symmetry~\cite{okuma2023non}.

\acknowledgments
N.C. acknowledges financial support from the Working Internships in Science and Engineering (WISE) fellowship from the German Academic Exchange Serivce (Deutscher Akademischer Austauschdienst, DAAD) as well as the Kishore Vaigyanik Protsahan Yojana (KVPY) fellowship from the Dept. of Science and Technology, Govt. of India.
A.G.M. acknowledges financial support from the Academy of Finland (Project 331094) and Jane and Aatos Erkko Foundation.
We acknowledge financial support from the German Research
Foundation (Deutsche Forschungsgemeinschaft, DFG)
under Germany’s Excellence Strategy through the W\"{u}rzburg-Dresden Cluster of Excellence on Complexity and Topology in Quantum Matter—ct.qmat (EXC 2147, Project No. 390858490).

\appendix

\begin{widetext}

\section{\label{app:Sylvester}Topological invariant for the non-Hermitian system}

Here we derive an alternate expression for the localizer index in Eq.~\eqref{eq:localizer} applied to doubled Hamiltonians of the form in Eq.~\eqref{eq:chiral_doubled}.
The position operator for the doubled system has the block form:
\begin{equation}
    \label{eq:doubled_position}
    \widetilde{X}=
    \begin{pmatrix}X & \mathbf{0}\\ \mathbf{0} &  X\end{pmatrix}
\end{equation}
$X=\rm{diag}(x_1-x_0,x_2-x_0,...x_N-x_0)$ is the position operator for the non-Hermitian system.
Substituting the block representations of the matrices in Eq.~\eqref{eq:localizer}, we get the invariant as
\begin{equation}
    \nu_L=\frac{1}{2}\rm{Sig}
    \begin{pmatrix}
    X& -iH\\
    iH^{\dagger} & -X\end{pmatrix}.
\end{equation}
The matrix in the argument is the localizer $L$, and its signature can be written in terms of the signature of its blocks.
We use the following identity for the block LDU decomposition:
\begin{equation}
    \label{eq:LDU}
    \begin{pmatrix} A & B
                    \\ C & D 
    \end{pmatrix}
    =
    \begin{pmatrix}\mathbb{I} & \mathbf{0}
                   \\CA^{-1}  & \mathbb{I}
    \end{pmatrix}
    \begin{pmatrix}  A  &  \mathbf{0}\\ 
                    \mathbf{0} & D-CA^{-1}B
    \end{pmatrix} 
    \begin{pmatrix}\mathbb{I} & A^{-1}B\\
                   \mathbf{0} &\mathbb{I}
    \end{pmatrix}
\end{equation}
Assuming that $X$ is invertible, meaning that we choose $x_0\neq x_j$ for all $j$, we apply this block LDU decomposition for $L$, resulting in:
\begin{equation}
\label{eq:congruence}
    L=SMS^{\dagger},
\end{equation}
where 
\begin{subequations}
\begin{equation}
S=\begin{pmatrix}\mathbb{I} & \mathbf{0}\\
    iH^{\dagger}X^{-1} & \mathbb{I}
    \end{pmatrix}
\end{equation}
and
\begin{equation}
M=\begin{pmatrix}
        X & \mathbf{0}\\
        \mathbf{0} &-X-H^{\dagger}X^{-1}H
    \end{pmatrix}
\end{equation}
\end{subequations}
Eq.~\eqref{eq:congruence} is called a congruence relation, which by Sylvester's law implies that ${\rm Sig}(L)={\rm Sig}(M)$ ~\cite{BUENO200763, Horn2012matrixanalysis, Horn2017linearalgebra}.

Since $M$ is block diagonal, its signature is simply the sum of the signatures of the diagonal blocks.
Therefore, the localizer index can be written in terms of the quantities for the non-Hermitian system as:
\begin{equation}
    \label{eq:localizer_nh}
    \nu_L=\frac{1}{2} {\rm Sig}(X)-\frac{1}{2}{\rm Sig}\left( X+H^{\dagger}X^{-1}H^{\pd} \right)
\end{equation}
Aside from providing an eight-fold speed-up for numerical computation, this also provides the expression for the invariant directly in terms of the non-Hermitian system and gives the same answer as the invariant calculated on the doubled system, as shown in Fig.~\ref{fig:NH_Index}.

\begin{figure}[t]
\includegraphics[width=0.3\columnwidth]{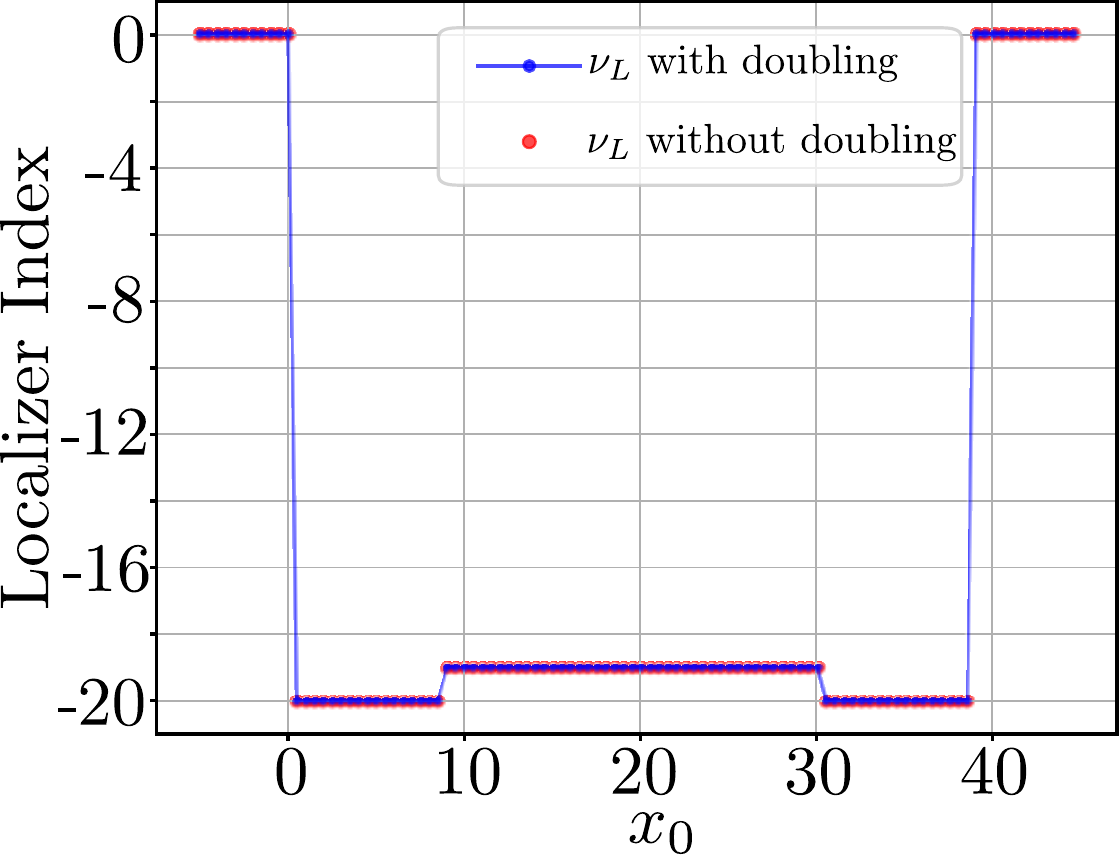}
\caption{
Localizer index calculated using Eq.~\eqref{eq:localizer} (blue) and Eq.~\eqref{eq:localizer_nh} (red). Both approaches give the same value for the invariant for the 40$\times$20 system with a dislocation with $B_y=1$, with a significantly faster computation for the non-Hermitian invariant in Eq.~\eqref{eq:localizer_nh}.
}
\label{fig:NH_Index}
\end{figure}

\section{Determining error bars}
\label{app:errorbar}
The choice of the error bars is made in accordance with the characteristics of the distribution of the corresponding variable.
Since the energy gap is a continuous variable which is bounded below by 0, we use the first and third quartiles of the distribution as the lower and upper error bounds.
This measure of the dispersion of the data is known as the interquartile range, and gives the interval where the middle half of data points are contained.

Since the topological index is always integer by definition, the corresponding distribution is discrete. 
We thus use the bootstrap method~\cite{press2007numerical}.
We create data sets of the same size as the original data set (100 disorder realizations for each $W$), by randomly resampling data points from the data set with replacement.
We calculate the mean for each resampled set, and use the interquartile range on the data set of means for each such resampled set.
For each disorder value, we apply the bootstrap procedure 10000 times and report the interquartile range as the lower and upper error bars.

\end{widetext}

\section*{References}
\bibliography{references}

\end{document}